\documentclass[preprint,5p,times,twocolumn]{elsarticle}

\usepackage{amssymb}
\usepackage{amsmath}
\usepackage{url}
\usepackage{booktabs}
\usepackage{threeparttable}
\usepackage{listings}
\journal{Pattern Recognition Letters}

\begin{document}

\begin{frontmatter}



\title{Function-based Labels for Complementary Recommendation: \\Definition, Annotation, and LLM-as-a-Judge}


\author[1]{Chihiro Yamasaki}
\author[1]{Kai Sugahara}
\author[1]{Yuma Nagi}
\author[1]{Kazushi Okamoto}

\affiliation[1]{organization={The University of Electro-Communications},
            addressline={1-5-1 Chofugaoka}, 
            city={Chofu},
            postcode={182-8585}, 
            state={Tokyo},
            country={Japan}}

\begin{abstract}
Complementary recommendations enhance the user experience by suggesting items that are frequently purchased together while serving different functions from the query item.
Inferring or evaluating whether two items have a complementary relationship requires complementary relationship labels; however, defining these labels is challenging because of the inherent ambiguity of such relationships.
Complementary labels based on user historical behavior logs attempt to capture these relationships, but often produce inconsistent and unreliable results.
Recent efforts have introduced large language models (LLMs) to infer these relationships.
However, these approaches provide a binary classification without a nuanced understanding of complementary relationships.
In this study, we address these challenges by introducing Function-Based Labels (FBLs), a novel definition of complementary relationships independent of user purchase logs and the opaque decision processes of LLMs.
We constructed a human-annotated FBLs dataset comprising 2,759 item pairs and demonstrated that it covered possible item relationships and minimized ambiguity.
We then evaluated whether machine learning methods using annotated FBLs could accurately infer labels for unseen item pairs, and whether LLM-generated complementary labels align with human perception.
Among machine learning methods, ModernBERT achieved the highest performance with a Macro-F1 of 0.911, demonstrating accuracy and robustness even under limited supervision.
For LLMs, GPT-4o-mini achieved high consistency (0.989) and classification accuracy (0.849) under the detailed FBL definition, while requiring only 1/842 the cost and 1/75 the time of human annotation.
Overall, our study presents FBLs as a clear definition of complementary relationships, enabling more accurate inferences and automated labeling of complementary recommendations.
\end{abstract}




\end{frontmatter}



\section{Introduction}\label{sec:introduction}

Item-to-item recommender systems suggest alternative items to users based on the relationships between the items.
Complementary recommendation is a type of item-to-item recommender system that focuses on two aspects of item relationships: complementary and substitute relationships; it recommends items that have a complementary relationship with a given item~\cite{2024mar_l.li}.
For instance, the relationship between smartphone A and another company's smartphone B is considered a substitute relationship because they serve the same purpose, whereas a protective film for a smartphone is regarded as complementary because it enhances its functionality.
By suggesting items based on these complementary relationships, the system assists users in item searches and contributes to increased sales on e-commerce sites by promoting co-purchasing behavior~\cite{2024mar_l.li,2021dec_n.entezari}.

Automatically identifying whether a given pair of items is complementary is necessary to make complementary recommendations.
However, defining complementary relationships is challenging because of the complex and diverse relationships among items~\cite{2019may_h.yu}.
For instance, a pair such as ``frying pan and saucepan'' can be considered complementary because they can be used together for cooking, but they are not always used simultaneously; therefore, they may not always be complementary.
This ambiguity in determining whether a pair is complementary leads to varied interpretations among individuals, making it labor-intensive and costly to manually extract and define clear ``complementary pairs'' from numerous item combinations.
McAuley et al. proposed defining ``frequently co-purchased relationships'' as complementarities and ``frequently co-viewed relationships'' as substitutes to construct explicit relationship labels, using co-occurrence counts in user purchases and viewing histories to label item pairs~\cite{2015aug_j.mcauley,2020oct_j.hao}.
Several previous studies have used these behavior-based labels constructed from such history-based labeling methods as labels for model training and evaluation.
Nevertheless, it has been noted that actual histories contain irregular user behaviors, resulting in numerous noisy labels that do not match the true relationships~\cite{2023sep_r.papso,2024oct_z.li,2024oct_k.sugahara}.

Recent studies~\cite{2023sep_r.papso,2024oct_z.li,2024feb_q.zhao} have attempted to mitigate noisy labels using Large Language Models (LLMs) by leveraging their vast knowledge and excellent reasoning capabilities to capture complementary semantic relationships.
These methods typically rely on the output of LLMs generated from prompts that ask binary questions, such as ``Is it a complementary relationship?'' or ``Is it a substitute relationship?'' without considering the detailed nuances of the item relationships.
Thus, similar to humans, LLMs may respond ambiguously when assessing item relationships, potentially leading to incorrect reasoning or explanation generation.
Moreover, determining whether LLMs can appropriately capture item relationships in the same way that humans perceive them has not been sufficiently evaluated.
Most existing studies have focused primarily on accuracy metrics~\cite{2023sep_r.papso,2024oct_z.li,2024feb_q.zhao}, and the evaluation of LLM-as-a-Judge for complementary relationships has thus far been limited to binary assessments by humans, such as whether the classification is complementary and whether the explanation is useful~\cite{2024oct_z.li}.
Owing to these limitations, uncertainties remain regarding whether the LLMs outputs align with human perceptions.

Consequently, this study aims to develop Function-Based Labels (FBLs) that explicitly capture item relationships based on their inherent functional characteristics, independent of historical behavioral data or the opaque decision-making processes of LLMs.
Our primary goal is to determine whether human-annotated FBLs can effectively describe functional relationships and whether machine learning (ML) models, including LLMs, can learn to infer these relationships in a manner that aligns with human judgments.
Specifically, we aim:
\begin{itemize}
    \item{To define FBLs that exclusively rely on the functional properties of items, avoiding confounding effects from historical user behavior or purchase logs.}
    \item{To construct a comprehensive FBL dataset by manually annotating item pairs, thereby establishing a robust ground truth for evaluating functional relationships.}
    \item{To investigate the capability of ML models to predict FBLs for unseen item pairs, and to examine how closely the inferences of LLMs align with human-annotated FBLs.}
\end{itemize}
This study raises three research questions:
\begin{description}
 \item{{\textbf{RQ1:}} Can the functional relationships clarify item relationships in complementary recommendations?}
 \item{{\textbf{RQ2:}} Can ML models trained on a small FBLs dataset accurately infer relationships for new item pairs?}
 \item{{\textbf{RQ3:}} Are the FBLs inferred by LLMs consistent with human-annotated FBLs?}
\end{description}
To answer these questions, for RQ1, we analyzed the distribution and clarity of the manually assigned FBLs.
For RQ2, we used the constructed FBLs dataset to train and evaluate the ML models and examine the mechanical identifiability of FBLs.
For RQ3, we designed prompts for LLMs to estimate relationships in FBLs from item information, and analyzed the consistency and agreement with manual annotations for both the proposed FBLs categories and the three traditional categories, thereby evaluating the usefulness of LLMs as annotators.
This study is an extension of a previous study~\cite{2024jan_k.sugahara}.

\section{Related Work}\label{sec:relatedWork}

\subsection{Complementary Recommendation}

The definition of ``complementary relationship'' in complementary recommendation varies.
A survey paper~\cite{2019may_h.yu} comprehensively summarizes complementary recommendations and states that complementary items are those that are different from the query item, but are sold together as accompanying items.
This definition serves as a broad criterion for determining whether an item is complementary; however, it contains several ambiguities, and no examples exist of more detailed definitions of complementary relationships.

Similar to general recommender systems, complementary recommendations involve two approaches: unsupervised and supervised learning~\cite{2024mar_l.li,2019may_h.yu}.
The main concept of unsupervised learning methods is to identify complementary items based on the co-occurrence frequency of co-purchases between items or between items and users, along with a threshold.
However, co-purchase data often contain both complementary and substitute items, making it difficult to accurately detect complementary items because of this noise.
To address this issue, Zheng et al.~\cite{2009jul_j.zheng} divided a series of browsing and purchasing behaviors into comparison and consideration stages, defining a complementary score that also considers substitute items.
In addition, methods that generate embedded representations from item and user co-occurrences~\cite{2018oct_m.wan} have been proposed.

By contrast, supervised learning methods can be interpreted as link prediction tasks between items.
In practice, the accurate labeling of complementary pairs with complex relationships requires annotators with deep domain knowledge and significant effort, and datasets recording complementary relationships do not exist~\cite{2024mar_l.li}.
In an initial attempt to label the relationships between items, McAuley et al. proposed a behavior-based labeling method using data collected from Amazon.com.
This method categorizes four types of item graphs based on user behavior: (1) users who viewed $x$ also viewed $y$, (2) users who viewed $x$ eventually purchased $y$, (3) users who purchased $x$ also purchased $y$, and (4) $x$ and $y$ are frequently purchased together.
They defined (1) and (2) as substitute relationships and (3) and (4) as complementary relationships~\cite{2015aug_j.mcauley}.
These behavior-based labels have been widely used in numerous previous studies for training and evaluating models~\cite{2015aug_j.mcauley,2020oct_j.hao,2018feb_z.wang,2019jan_v.rakesh,2020jan_d.xu,2021apr_m.angelovska,2021dec_l.ma}.
To address the cold start problem, content-based methods that capture complementary relationships based on item text and image information have been studied.
LDA-based approaches have also been explored for generating item embeddings from reviews; however, their limitations in handling short texts have led to the development of deep generative models using variational autoencoders~\cite{2019jan_v.rakesh}.
Furthermore, efforts have been made to generate item embeddings with higher expressiveness using the available item information (title, description, brand, and images)~\cite{2023sep_r.papso,2021apr_m.angelovska}.

\subsection{Addressing Noisy Labels}

Complementary pairs estimated based on user viewing and purchasing behavior, as well as the constructed training labels, do not always accurately reflect true complementary relationships and may contain numerous noisy labels~\cite{2023sep_r.papso,2024oct_z.li,2020jan_d.xu,2021dec_l.ma}.
The main causes of this issue include noises from the co-purchase of popular items and irregular user behavior.
In the former case, the high frequency of co-purchases with popular items bought by several users increases the risk of incorrectly estimating non-complementary item pairs as complementary pairs~\cite{2021dec_l.ma}.
In the latter case, irregular user behavior introduces significant bias in estimating complementary pairs, which is particularly problematic for e-commerce sites and item pairs with limited behavioral histories~\cite{2023sep_r.papso}.
Such real-world label information inevitably contains considerable noise, which can severely impair model performance~\cite{2024jun_y.zhang}.

Several studies have focused on reducing the noise in behavior-based labels to create more refined labels.
Hao et al.~\cite{2020oct_j.hao} proposed a strategy to obtain more accurate complementary relationship labels by considering the difference between a set of complementary relationship pairs and a set of substitute relationship pairs from McAuley et al.'s four types of item graphs~\cite{2015aug_j.mcauley}.
In addition, methods have been proposed to effectively extract item relationships from noisy browsing, purchasing, and search behaviors and item information by introducing a self-attention mechanism~\cite{2020jan_d.xu}.
Another approach models the co-purchase frequency of each item as a Gaussian distribution, separating co-purchases based on complementary relationships from those due to noise, to estimate more genuine complementary relationships~\cite{2021dec_l.ma}.

Recent studies have used LLMs, which demonstrate high performance in zero-shot estimation tasks, to infer complex complementary relationships~\cite{2023sep_r.papso,2024oct_z.li,2024feb_q.zhao}.
These studies have demonstrated the usefulness of LLMs in extracting complementary relationships for item pairs that behavior-based labels cannot capture, such as long-tail items~\cite{2023sep_r.papso,2024feb_q.zhao}, and in generating explanations of relationships using their advanced language generation capabilities~\cite{2024oct_z.li}.
In these methods, prompts are provided to LLMs to determine whether the target item pair has a ``complementary relationship'' or a ``substitute relationship,'' and the outputs are used for knowledge expansion~\cite{2023sep_r.papso,2024oct_z.li} and constructing item graphs~\cite{2024feb_q.zhao}.
However, the extent to which the relationship estimations of LLMs align with human judgment has not been sufficiently verified.
Previous studies have often focused more on improving the accuracy of recommendation tasks than on the accuracy of the relationship estimations of LLMs.
Evaluations of how LLMs determine relationships and their consistency with human perceptions are limited.
Moreover, several methods estimate relationships by asking LLMs binary questions such as ``Is it a complementary relationship?'' or ``Is it a substitute relationship?''
However, this approach makes it difficult to consider the strength of the relationships or subtle semantic nuances.
Consequently, the inference results may be used for knowledge expansion or item graph construction without being sufficiently accurate.
Li et al.~\cite{2024oct_z.li} introduced the concept of ``consistent complementary relationships,'' considering the asymmetry of item relationships, and evaluated the classification accuracy of LLMs using manually labeled correct data, but the evaluation was limited to 200 pairs.
\section{Function-based Labels}\label{sec:defLabel}

\subsection{Definition}
This study focuses on the functional relationships between items to clarify ambiguous complementary relationships, excluding influences based on viewing or purchase relationships.
Specifically, through discussions with e-commerce site practitioners, we define nine types of relationship labels based on item functions, known as FBLs, as follows:
\begin{enumerate}
    \item[(A)] Items $x$ and $y$ have the same function and usage.
    \item[(B-1)] Item $x$ can be replenished with item $y$.
    \item[(B-2)] Item $y$ can be replenished with item $x$.
    \item[(C-1)] Item $x$ and $y$ must be combined to be usable.
    \item[(C-2)] When combined with item $y$, item $x$ becomes more useful.
    \item[(C-3)] When combined with item $x$, item $y$ becomes more useful.
    \item[(C-4)] Combining $x$ and $y$ makes them more useful.
    \item[(D)] Items $x$ and $y$ have no relationship.
    \item[(E)] Items $x$ and $y$ seem to have a relationship, but it is difficult to verbalize.
\end{enumerate}

Several pre-tests were performed to finalize the labels.
In each pre-test, we (1) extracted 100 sample pairs, (2) assigned relationship labels, (3) evaluated their validity and coverage, and (4) iteratively revised the label definitions as necessary.
During this process, we focused on functional aspects to ensure that the relationships were defined systematically and comprehensively.
The three traditional categories of relationships (substitute, complementary, and unrelated) are mapped to the new detailed labels as follows: the substitution relationship corresponds to Label A, complementary relationship corresponds to Labels B-* and C-*, and unrelated (neither substitution nor complementary relationship) corresponds to Label D.
In complementary relationships, a sequential purchase order where one item becomes necessary after purchasing another is assumed to exist.
Therefore, we define Label B-* for ``replenishment'' to distinguish it from Label C-*.
For instance, the correction tape body (Item x) and refill tape (Item y) illustrate this.
Additionally, Label C-* captures cases in which one of the paired items can (or cannot) be used independently for its intended function.
For instance, a broom (Item x) can be used alone for cleaning, but combining it with a dustpan (Item y) makes cleaning more convenient, corresponding to Label C-2.
Another example is an insert cup (Item x) and its holder (Item y), which cannot be used independently and only become functional when combined; thus, it should be assigned Label C-1.
Finally, Label E was used for complementary relationships that could not be clearly defined by the authors.
As mentioned earlier, complementary relationships have complex structures; therefore, the pairs assigned to Label E should be used as the basis for future discussions and definitions of the relationships.

These relationships were intended for use in the annotation process, as described later.
Labels with clearer relationship definitions were placed earlier in the sequence from Label A to Label E to ensure the efficiency and reliability of annotation.
Annotators use a flowchart to determine whether an item pair fits each definition in order.
Based on insights from the pre-tests, we subdivided complementary relationships into finer categories (Labels B-* and C-*) to better reflect functional distinctions and to facilitate annotation efficiency by reducing annotator burden.

\subsection{Human Annotation}

\subsubsection{Annotation Setup}
We prioritized annotator familiarity and item relationship complexity to ensure accurate and effective label collection, selecting pairs from the well-known everyday categories ``Office Supplies/Stationery'' and ``Household Goods/Kitchenware.''
Random sampling from all item pairs would result in most pairs being classified as ``unrelated,'' making it difficult to obtain sufficient pairs with substitute and complementary relationships.
Conversely, sampling from item pairs with high co-purchase counts could lead to a bias toward certain item categories owing to differences in the number of purchased items, thereby losing the diversity of the annotation target pairs.
Therefore, in this experiment, we used the D'Hondt method~\cite{2013jun_n.bormann}, based on co-purchase counts between item categories, to ensure diversity and sample a certain number of pairs with various relationships.
In the D'Hondt allocation method, the co-purchase count for each major category pair is divided by natural numbers ($1, 2, 3, \cdots$), and the sampling numbers are allocated in order from the largest quotient.
We then sample the top co-purchased item pairs according to the allocated sampling numbers for each major category pair.
Ultimately, we selected 2,000 pairs using the D'Hondt method and added 800 pairs randomly selected from those already empirically identified as complementary by field practitioners, resulting in 2,800 pairs for annotation.
We asked 18 annotators (7 undergraduate computer science students, 9 graduate computer science students, and 2 faculty staffs) to assign unique labels based on the item titles and web pages of the item pairs, with each annotator handling 400 or 500 pairs and collecting 8,400 responses.
The final label was determined by majority vote among the three annotators, and only labels selected by at least two annotators were adopted.

\subsubsection{Annotation Results}
\begin{figure}[t]
    \centering
    \includegraphics[width=0.8\linewidth]{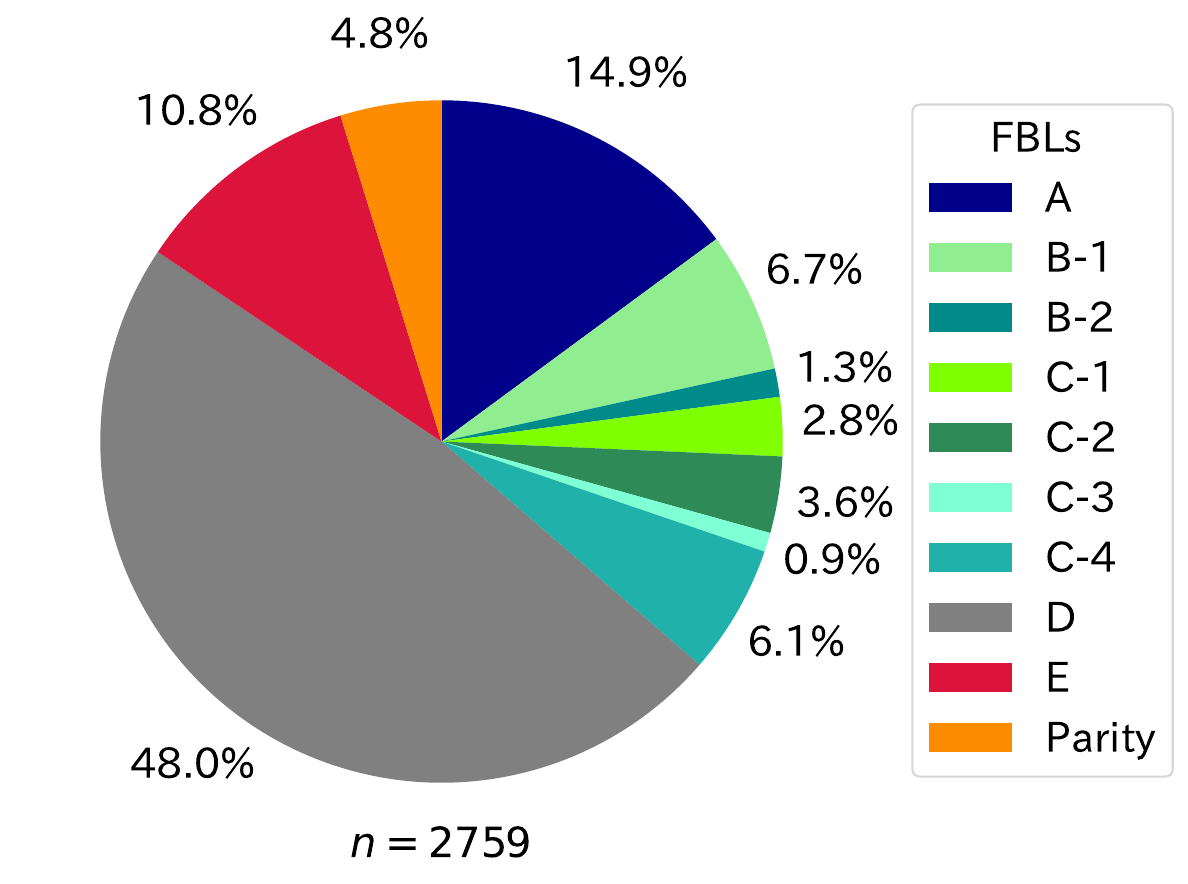}
    \caption{Label distribution in the human-annotated FBLs $\mathfrak{F}_{\text{Human}}$} \label{fig:labelRate_Human}
\end{figure}

Human annotation required on average 16 JPY and 50 seconds per item pair.
Of the 2,800 pairs collected through annotation, 41 that did not fit the FBLs definitions were excluded as invalid responses.
This resulted in a human-annotated FBL dataset, $\mathfrak{F}_{\text{Human}}$, comprising 2,759 pairs.
The label distribution in $\mathfrak{F}_{\text{Human}}$ is shown in Figure~\ref{fig:labelRate_Human}.
Label A, which corresponds to substitute relationships, accounts for 14.9\%.
Labels B-* and C-*, which correspond to complementary relationships, comprised 21.4\%.
Label D, which represents unrelated relationships, was the most common at 48.0\%.
Label E, which indicates relationships that could not be defined, was relatively low at 10.8\%, suggesting that the proposed function-based relationship definitions covered most relationships.
Notably, most pairs labeled as Label E were qualitatively observed to exhibit indirect complementary through a third item (e.g., an envelope and a pen refill through a pen).
Such relationships cannot be adequately represented by binary relations alone and require consideration of ternary or higher-order relations.
From the viewpoint of binary relation modeling, assigning these cases to Label E can therefore be regarded as an appropriate classification.

\begin{figure}[t]
    \centering
    \includegraphics[width=\linewidth]{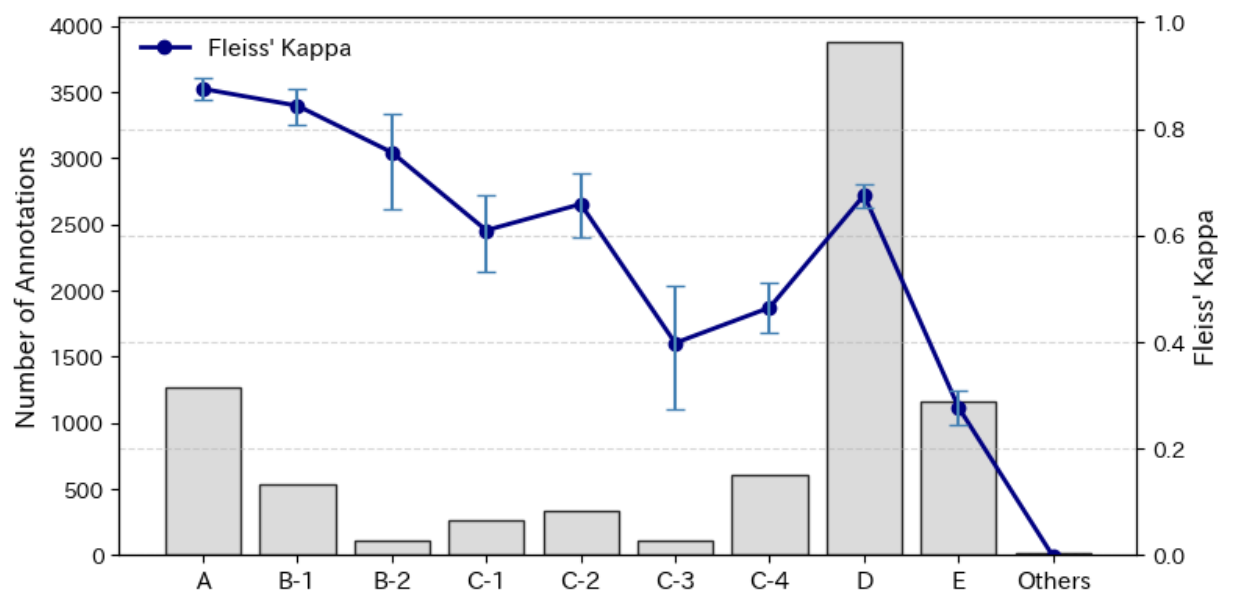}
    \caption{Fleiss' Kappa with 95\% CI in the human-annotated FBLs $\mathfrak{F}_{\text{Human}}$} \label{fig:fleiss_kappa}
\end{figure}

Regarding annotation reliability, the proportion of parity cases was low at 4.8\%.
Fleiss' Kappa~\cite{1971xxx_j.l.fleiss} further yielded an overall value of 0.63 (moderate agreement), and Figure~\ref{fig:fleiss_kappa} shows the Kappa values with 95\% confidence intervals (95\% CI) for each label.
Agreement was relatively high for Labels A and B-*, whereas lower values were observed for some fine-grained complementary categories (Labels C-3 and C-4) and for Label E.
These results suggest that the proposed FBLs provided sufficiently clear and consistent definitions for human annotation, while certain categories still posed greater challenges.
For subsequent evaluation experiments, we used $\mathfrak{F}'_{\text{Human}}$, which comprised 2,625 pairs, excluding the contested pairs from $\mathfrak{F}_{\text{Human}}$.
The FBLs dataset developed in this study is publicly available on GitHub\footnote{\url{https://github.com/okamoto-lab/fbl_dataset}}.
The dataset contains 2,800 item pairs with their corresponding relationship labels, including web links to each item in the pairs, to support further research on complementary recommender systems.

\section{Evaluation Experiment in Machine Learning Models}\label{sec:experimentalML}

\subsection{Settings for Machine Learning Models}
As an initial step, our study aims to predict whether two items are complementary (compl.), substitute (subst.), or unrelated (unrel.).
We therefore evaluated the classification accuracy of ML models using $\mathfrak{F}'_{\text{Human}}$ in this three-class setting based on item attribute features of item pairs.
The nine-class labels, where directionality was introduced only in annotations, were mapped to the three classes as follows:
Label A was mapped to substitute, Labels B-* and C-* were mapped to complementary,
and Labels D and E were mapped to unrelated.

We employed four traditional models: Logistic Regression (LR), LightGBM (LGBM), Support Vector Machine (SVM), $k$-Nearest Neighbors ($k$NN), and two neural models: TextCNN~\cite{2014oct_y.kim} and ModernBERT-Ja-70m~\cite{2025apr_i.sugiura}~\footnote{\url{https://huggingface.co/sbintuitions/modernbert-ja-70m}}.
For both neural models, we used the tokenizer of ModernBERT-Ja-70m.
We adopted double cross-validation~\cite{2009feb_p.filzmoser} with 5-fold cross-validation on the outer layer (outer test set) and stratified 5-fold cross-validation on the inner layer (inner validation set).
For the traditional models, the inner validation set was used for parameter tuning with Optuna~\cite{2019xxx_t.akiba} to prevent overfitting and assess generalizability.
For the neural models, the inner set was used to stabilize training and for early stopping.
As the evaluation metric and tuning objective, we employed Macro-F1 across the three classes {complementary, substitute, and unrelated}, in order to account for class imbalance.
Macro-F1 was defined as the mean of class-wise F1 scores:
$$
\text{Macro-F1} = \frac{1}{3} \sum_{i \in \{\text{compl.}, \text{subst.}, \text{unrel.}\}} \text{F1}_i
$$
where $\text{F1}_i = 2 \cdot \text{Precision}_i \cdot \text{Recall}_i / (\text{Precision}_i + \text{Recall}_i)$.

As input to the traditional models, we construct a 424-dimensional feature vector $\boldsymbol{x_i = [x^{sim}_i, x^{mch}_i, x^{cate}_i]}$, combining three types of features:
(i) Similarity features: cosine similarities between multilingual-e5-large\footnote{\url{https://huggingface.co/intfloat/multilingual-e5-large}} embeddings of item information (title, description, specification, category),
(ii) Match flag features: binary indicators of exact matches for major category, subcategory, manufacturer, and brand, and
(iii) Category match features: binary indicators showing whether item pairs match in each element of the major and subcategory classifications.
For the neural models, the input was constructed as a single text sequence by concatenating the attributes of each item (title, description, category, maker, and brand) in this order, separated by commas.
A delimiter is then placed between the sequences of item $x$ and item $y$ to distinguish the two items.

\subsection{Comparison of Classification Accuracy by ML Models}
\begin{figure}[t]
    \centering
    \includegraphics[width=\linewidth]{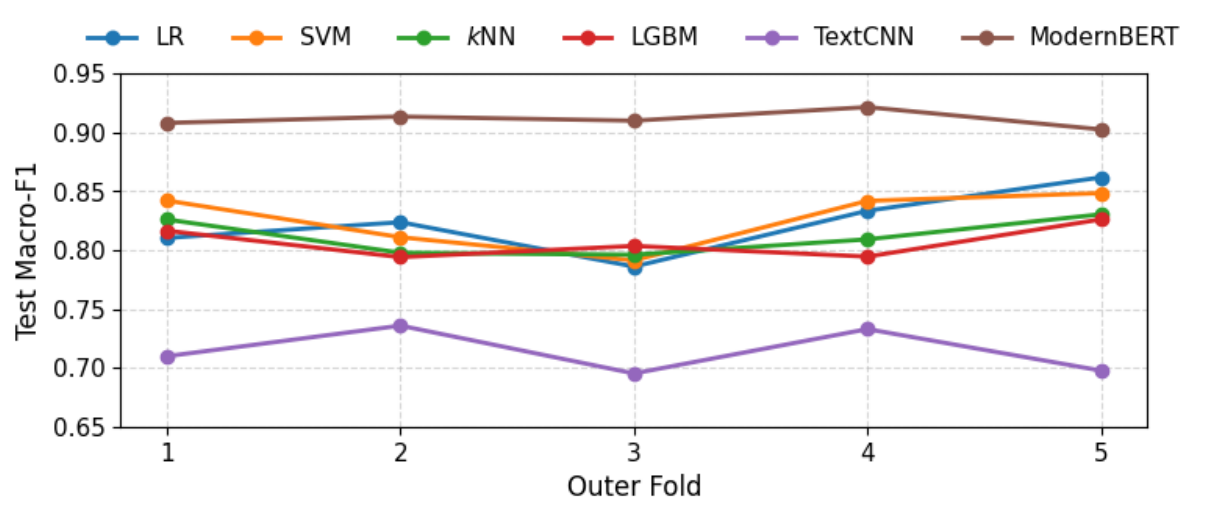}
    \vspace{-7mm}
    \caption{Macro-F1 on the test set across outer folds for each model}
    \label{fig:ml_result_outerfolds}
\end{figure}

Table~\ref{tab:ML_result} summarizes the classification accuracy of all models in the double cross-validation.
Among the traditional models, LR and SVM achieved the highest Macro-F1 (around 0.82), while $k$NN and LGBM were slightly lower (about 0.80).
The relatively lower performance of LGBM may be explained by its feature selection behavior, which made more limited use of features compared to LR (43 selected features for LGBM vs. 151 for LR out of 424).
TextCNN performed clearly worse (around 0.71), whereas ModernBERT markedly outperformed all models, reaching 0.898 (inner validation set) and 0.911 (outer test set).
As shown in Figure~\ref{fig:ml_result_outerfolds}, certain folds yielded higher the outer test accuracies, likely reflecting fold-level variability with some folds containing easier to classify pairs for each models.
Such variability was evident in most models, whereas ModernBERT achieved the highest accuracy with comparatively small fluctuation across folds, highlighting its stability and performance.

\begin{table}[t!]
  \centering
  \footnotesize
  \caption{Comparison of Macro-F1 by ML models (mean $\pm$ standard deviation across 5-fold double cross-validation)}
  \label{tab:ML_result}
  \begin{tabular}{lcc}
      \hline
      & Validation set & Test set \\
      \hline
      LR & 0.818 ($\pm$ 0.007) & 0.823 ($\pm$ 0.028) \\
      SVM & 0.818 ($\pm$ 0.019) & 0.826 ($\pm$ 0.025) \\
      $k$NN & 0.794 ($\pm$ 0.011) & 0.811 ($\pm$ 0.015) \\
      LGBM & 0.800 ($\pm$ 0.003) & 0.802 ($\pm$ 0.012) \\
      ModernBERT & 0.898 ($\pm$ 0.014) & 0.911 ($\pm$ 0.006) \\
      TextCNN & 0.711 ($\pm$ 0.024) & 0.714 ($\pm$ 0.017) \\
      \hline
  \end{tabular}
\end{table}

Overall, these results indicate that while traditional models such as LR and SVM already achieve solid performance under limited supervision, advanced neural approaches such as ModernBERT provide even higher accuracy and robustness.
Robustness to small datasets is a critical challenge when working with manually labeled data, and our findings offer useful insights for developing methods that exploit the FBL dataset with its limited number of labels.

\section{Evaluation Experiment for LLMs}\label{sec:experimentalLLM}

\subsection{Annotation Settings for LLMs}
We evaluated the usefulness of the LLMs as annotators by comparing their annotation results with $\mathfrak{F}'_{\text{Human}}$.
In this experiment, we examine the results of two LLMs: gpt-4o-mini (GPT)~\cite{2024oct_openai} and an 8-bit quantized model of Llama-3.1-70B~\cite{2024jun_a.grattafiori}, which is a Japanese fine-tuned model\footnote{https://huggingface.co/cyberagent/Llama-3.1-70B-Japanese-Instruct-2407}.
For the LLM parameter settings, we set the temperature, which adjusts the randomness of responses, to 0.6.
This setting is based on the analysis by Pangakis et al.~\cite{2023may_n.pangakis}, which confirmed that setting the temperature to 0.6 or higher and using majority voting from multiple (at least seven times) outputs improves the labeling accuracy of LLMs.
Additionally, for Llama, we set mirostat\_tau, a parameter that adjusts response consistency, to 1, which is lower than the default value of 5, to suppress unnecessary verbose explanations in the labeling tasks.
The LLMs were instructed to output a unique label based on the titles, descriptions, category names, manufacturer names, and brand names of the input item pairs, according to the following two label systems:
\begin{enumerate}
  \item{Nine relationships focused on item functionality (9-cls)}
  \item{Three relationships considered in previous studies~\cite{2023sep_r.papso,2024oct_z.li,2024feb_q.zhao}: complementary, substitute, and unrelated (3-cls)}
\end{enumerate}
In each label system, a unique label is determined by majority voting from the output label list $\hat{\boldsymbol{y}}$ obtained from the seven inference results.
The experiments were conducted using Japanese prompts, and the English version of the 9-cls sample prompt is provided in \ref{app:prompt-9cls} to ensure reproducibility.

\begin{table}[t]
  \centering
  \footnotesize
  \caption{Comparison of Consistency Score by LLMs (3-cls: three traditional classes, 9-cls: nine FBL classes; majority voting from 7 outputs per item pair)}
  \label{tab:consistency_score}
  \begin{tabular}{lcc|cc}
    \toprule
    & \multicolumn{2}{c|}{Llama} & \multicolumn{2}{c}{GPT} \\
    \cmidrule(lr){2-3} \cmidrule(lr){4-5}
    & w/ 3-cls & w/ 9-cls & w/ 3-cls & w/ 9-cls \\
    \midrule
    Compl. & 0.936 & \textbf{0.939} & \textbf{0.995} & 0.985 \\
    Subst. & 0.841 & \textbf{0.876} & 0.959 & \textbf{0.976} \\
    Unrel. & 0.969 & \textbf{0.979} & 0.993 & \textbf{0.994} \\
    All    & 0.942 & \textbf{0.954} & 0.988 & \textbf{0.989} \\
    \bottomrule
  \end{tabular}
\end{table}

\subsection{Evaluation Metrics}
Two evaluation metrics were used to evaluate the annotation results of the LLMs.
In addition, the nine detailed relationship labels were consolidated into three categories for evaluation: Label A as a substitute, Labels B-* and C-* as complementary, and Labels D and E as unrelated.

\paragraph{1. Consistency Score}
We adopted Consistency Score proposed by Pangakis et al.~\cite{2023may_n.pangakis} to measure the consistency of the LLM outputs:
$$
\text{Consistency Score} = \frac{1}{\mid \hat{\boldsymbol{y}} \mid} \sum_{\hat{y} \in \hat{\boldsymbol{y}}} \left( \hat{y} = \hat{\boldsymbol{y}}_{mode} \right).
$$
Here, $\hat{\boldsymbol{y}}_{\text{mode}}$ is the most frequent label in the output label list $\hat{\boldsymbol{y}}$, and the average Consistency Score across all pairs serves as the final evaluation metric.
This metric assesses the agreement of multiple label outputs for each pair, indicating that the closer the score is to 1, the more clearly the LLM recognizes the item relationships and consistently outputs the same label.

\paragraph{2. Macro-F1}
As a metric to measure the agreement between the LLM-generated and human-annotated labels $\mathfrak{F}'_{\text{Human}}$, we used Macro-F1 for complementary, substitute, and unrelated labels, with human labels as the ground truth.

\subsection{Classification Accuracy of LLMs}
Tables~\ref{tab:consistency_score} and ~\ref{tab:macro_f1} summarize the LLM results for each label system.
In terms of Consistency Score, labeling with the 9-cls setting improved overall consistency compared with the 3-cls setting, achieving a high consistency of 0.989 with GPT w/ 9-cls.
For Llama, although its consistency was generally lower than that of GPT, the 9-cls setting contributed to a 1.2\% improvement in consistency.
In the substitute relationship, the 9-cls setting led to a notable consistency increase of 4.1\% for Llama and 1.7\% for GPT.
These results suggest that defining detailed relationships based on item functionality allows LLMs to understand item relationships more clearly, thus providing positive evidence for RQ1.

Similarly, in Macro-F1, the 9-cls setting achieved higher scores than the 3-cls setting did.
GPT w/ 9-cls achieved the best result with 0.849, followed by Llama w/ 9-cls at 0.822, achieving classification accuracy exceeding LR (0.823) without requiring labeled data.
Within the 9-cls setting, improvements in scores for complementary and substitute relationships were notable, with GPT achieving 0.797 for complementary and 0.816 for substitute relationships, exhibiting significant improvements over the 3-cls setting, with scores of 0.691 and 0.581, respectively.
These results indicate that a detailed label system based on item functionality enables LLMs to grasp the relationships between items more accurately, suggesting that LLMs can function effectively as annotators mimicking human perception.
Moreover, GPT-4o-mini required only 0.019 JPY and 0.67 seconds per pair, corresponding to 1/842 the cost and 1/75 the time of human annotation.
It maintained Macro-F1 of 0.849, demonstrating that such accuracy can be achieved at dramatically lower cost and latency.

\begin{table}[t]
  \centering
  \footnotesize
  \caption{Comparison of Macro-F1 by LLMs (3-cls: three traditional classes, 9-cls: nine FBL classes; majority voting from 7 outputs per item pair)}
  \label{tab:macro_f1}
  \begin{tabular}{lcc|cc|cc}
    \toprule
    & \multicolumn{2}{c|}{Llama} & \multicolumn{2}{c|}{GPT} & \multicolumn{2}{c}{ML (baseline)} \\
    \cmidrule(lr){2-3} \cmidrule(lr){4-5} \cmidrule(lr){6-7}
    & w/ 3-cls & w/ 9-cls & w/ 3-cls & w/ 9-cls & LR & Modern \\
    & & & & & & BERT \\
    \midrule
    Compl. & 0.688 & \textbf{0.782} & 0.691 & \textbf{0.797} & 0.705 & \textbf{0.866} \\
    Subst. & 0.519 & \textbf{0.758} & 0.581 & \textbf{0.816} & 0.837 & \textbf{0.905} \\
    Unrel.  & 0.901 & \textbf{0.924} & 0.914 & \textbf{0.933}  & 0.926 & \textbf{0.961} \\
    All    & 0.703 & \textbf{0.822} & 0.729 & \textbf{0.849} & 0.823 & \textbf{0.911} \\
    \bottomrule
  \end{tabular}
\end{table}

\subsection{Comparison between Human and LLMs FBLs}
\begin{figure}[t]
    \centering
    \includegraphics[width=.85\linewidth]{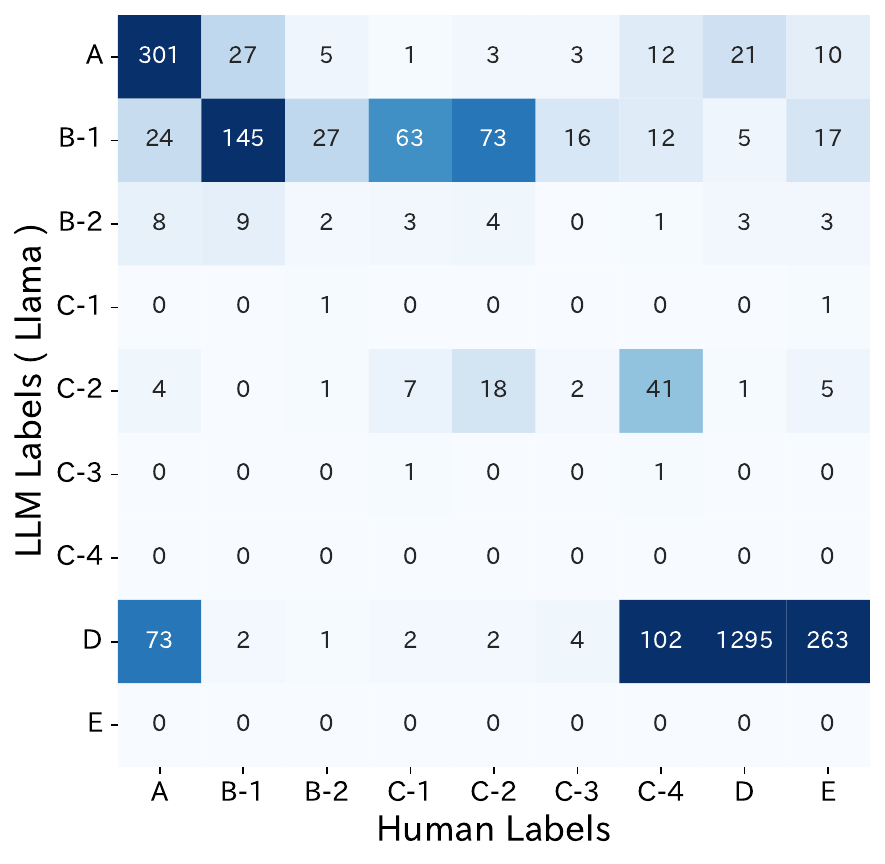}
    \caption{Correspondence between Llama-generated and human-annotated labels (counts, not percentages; color scale capped at 100 for visualization)}
    \label{fig:ConfusionMatrix_Llama}
\end{figure}

\begin{figure}[t]
    \centering
    \includegraphics[width=.85\linewidth]{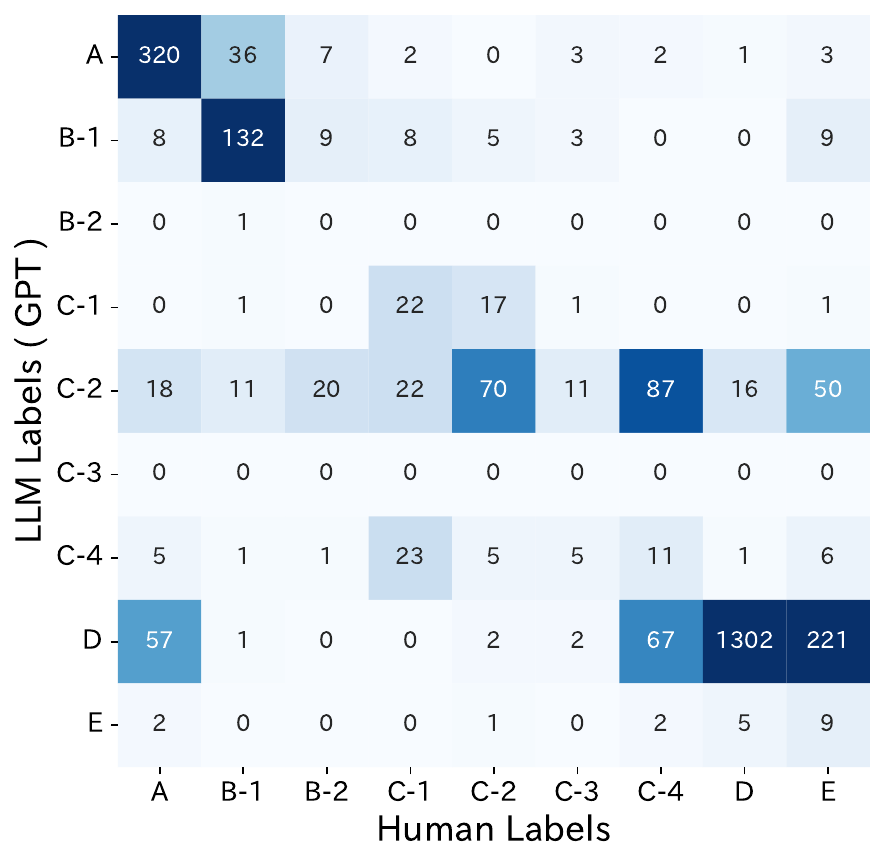}
    \caption{Correspondence between GPT-generated and human-annotated labels (counts, not percentages; color scale capped at 100 for visualization)}
    \label{fig:ConfusionMatrix_GPT}
\end{figure}

Figures~\ref{fig:ConfusionMatrix_Llama} and \ref{fig:ConfusionMatrix_GPT} show the confusion matrices comparing human-annotated FBLs with those generated by Llama and GPT in the 9-cls setting.
In Llama, outputs for complementary relationships are concentrated on Label B-1 (``Item $x$ can be replenished with item $y$'') indicating a relatively good classification of complementary relationships, but it struggles with distinguishing more complex relationships and their directions, such as replenishment and combination.
In contrast, GPT shows a reduced tendency for localization in complementary relationships, suggesting that it can capture more complex item relationships than Llama.
Both LLMs also tend to classify item pairs that humans labeled as ``A. Items $x$ and $y$ have the same function and usage.'' and ``C-4. Combining $x$ and $y$ makes them more useful.'' as ``D. Items x and y have no relationship.''
Examples of the former include pairs, such as envelopes of different sizes or pens with different ink properties, including mechanical pencils and oil-based pens, which are difficult for humans to judge as substitutes.
Examples of the latter include pairs such as sponges and bleach or disposable chopsticks and soup cups, which can be used together in the same scene (kitchen or dining), but do not necessarily have to be used, indicating a weak relationship.
Fleiss' Kappa in human annotator agreement was also low (approximately 0.4) for Labels C-3 and C-4, suggesting that the relationship definitions for these labels were less robust and likely contributed to the difficulty of correct labeling by LLMs.
The analysis of these item pair trends suggests that LLMs tend to diverge from human judgment for item pairs with ambiguous relationships or differences, functioning as stricter annotators than humans by classifying such pairs as unrelated.

\section{Conclusion}\label{sec:Conclusion}

In this study, we propose FBLs that focus on the functional relationships between items as a new definition of complementary relationships independent of user behavior history.
We constructed a human-annotated dataset based on the FBLs relationship definitions and evaluated the classification accuracy of these relationships using ML models and LLMs.
The label distribution results of the human-annotated FBL dataset clarified the existing relationships in complementary recommendations from a functional perspective.
Additionally, labeling by LLMs based on FBLs improved the response consistency, suggesting that FBLs function as a clear definition of complementary relationships (RQ1).
In the evaluation experiment of ML models using FBLs, LR and SVM achieved an average Macro-F1 of approximately 0.82 in both validation and test sets, demonstrating high classification accuracy (RQ2).
In the evaluation experiment of relationship estimation by LLMs, FBLs assigned by LLMs achieved a high Macro-F1 of up to 0.849 compared with human-annotated FBLs, suggesting that LLMs can function effectively as annotators mimicking human perception when using FBLs (RQ3).

While these findings demonstrate the promise of FBLs, several limitations remain to be addressed in future work:
\begin{itemize}
  \item{The current dataset is limited in size and category coverage; larger-scale annotation with additional annotators and diverse item domains will be necessary to confirm reproducibility and generalizability.}
  \item{Some fine-grained definitions, such as Labels C-3 and C-4, showed relatively low inter-annotator agreement (Kappa $\thickapprox$ 0.4), and Label E (10.8\%) captured cases that could not be clearly defined, indicating the need for refining and subdividing functional relationship definitions.}
  \item{The present study was conducted in Japanese; extending the dataset and experiments to other languages will allow us to consider cultural and linguistic influences.}
  \item{Expanding the dataset to include long-tail items will enable examination of domain shifts that naturally occur in real-world e-commerce.}
\end{itemize}

Although this study confirmed the usefulness of LLMs as annotators, applying LLMs to a vast number of item pairs incurs high costs in terms of time and money.
Therefore, to reduce these costs, the development of automatic pair classification using ML models is required.
In particular, active learning methods that utilize teacher labels obtained from LLMs~\cite{2023dec_r.xiao,2024sep_n.kholodna} are considered promising approaches and directions for future research.
Building on this perspective, it is also important to recognize that LLMs may not always possess sufficient knowledge to provide highly reliable labels.
Therefore, exploring active learning paradigms that explicitly account for uncertain labeling knowledge~\cite{2014xxx_m.fang} could further enable more robust and trustworthy utilization of LLM-generated FBLs.

\section*{Funding}
This work was supported by ASKUL Corporation and JSPS KAKENHI Grant Numbers JP23K21724 and JP24K21410.

\bibliographystyle{elsarticle-num}
\bibliography{references}

\appendix
\section{LLM Prompts for 9-class (English Version)}\label{app:prompt-9cls}

The prompt used for the 9-class annotation is shown below.

\begin{lstlisting}
Input: You are a highly capable assistant that assigns a unique label to the relationship between #ItemA and #ItemB by strictly following the #Procedure. In your output, do not include anything other than the label information in the format "{'label': <corresponding option number>}".

# Procedure
(1) Based on the detailed information of #Item A and #Item B, understand how each item is used.
(2) Go through the options in the #Flowchart (1) to (9) in order, and decide which option best matches the relationship between #ItemA and #ItemB.
(3) Output the corresponding option number strictly in the JSON format defined in #Output Format. Do not add any explanation or extra text.

# Flowchart
(1) Items A and B have the same function and usage.
(2) Item A can be replenished with item B.
(3) Item B can be replenished with item A.
(4) Item A and By must be combined to be usable.
(5) When combined with item B, item A becomes more useful.
(6) When combined with item A, item B becomes more useful.
(7) Combining A and B makes them more useful.
(8) Items A and B have no relationship.
(9) Items A and B seem to have a relationship, but it is difficult to verbalize.
(10) Others.

# Item A
{Input Item A Descriptions}

# Item B
{Input Item B Descriptions}

# Output Format
{'label': <corresponding option number>}

Output:
\end{lstlisting}

\end{document}